\input{aipcheck}

\documentclass[
    ,final            
  ]
  {aipproc}

\layoutstyle{8x11single}

\usepackage{amsmath}
\begin{document}

\title{Study of the $D^*\rho$ system using QCD sum rules}


\author{A.~Mart\'inez~Torres}{
  address={Instituto de F\'isica, Universidade de S\~ao Paulo, C. P. 66318, 05389-970 S\~ao Paulo, SP, Brazil.}
}
\author{K. P. Khemchandani}{
  address={Instituto de F\'isica, Universidade de S\~ao Paulo, C. P. 66318, 05389-970 S\~ao Paulo, SP, Brazil.}
}

\author{M. Nielsen}{
  address={Instituto de F\'isica, Universidade de S\~ao Paulo, C. P. 66318, 05389-970 S\~ao Paulo, SP, Brazil.}
}

\author{F. S. Navarra}{
  address={Instituto de F\'isica, Universidade de S\~ao Paulo, C. P. 66318, 05389-970 S\~ao Paulo, SP, Brazil.}
}

\author{E. Oset}{
  address={Departamento de F\'isica Te\'orica and IFIC, Centro Mixto Universidad de Valencia-CSIC, Institutos de Investigaci\'on de Paterna, Apartado 22085, 46071 Valencia, Spain.}
}
\begin{abstract}
In this talk I present a study of the $D^* \rho$ system made by using the method of QCD sum rules. Considering isospin and spin projectors, we investigate
the different  configurations and obtain three $D^*$ mesons with isospin $I=1/2$, spin $S=0$, $1$, $2$ and with 
masses $2500\pm 67$ MeV, $2523\pm60$ MeV, and $2439\pm119$ MeV, respectively. The last state can be related to $D^*_2(2460)$ (spin 2) listed by the Particle Data Group, while one of the first two might be associated with $D^*(2640)$, whose spin-parity is unknown. In the case of $I=3/2$ we also find evidences of three states with spin 0, 1 and 2, respectively,
with masses $2467\pm82$ MeV, $2420\pm128$ MeV, and $2550\pm56$ MeV. 
\end{abstract}

\maketitle


\section{Introduction}

Since the discovery of the $D$ and $D^*$ mesons~\cite{markI}, the interest in the charmed meson spectroscopy has grown exponentially and
experimental facilities like Cleo, Belle, BaBar, etc., keep bringing more relevant information about the excited charmed meson states~\cite{cleo,belle,babar}.

However, in spite of all the experimental and theoretical efforts, a rather scarce information is available on the properties of the charmed meson resonances like spin, isospin, parity, etc. 
The most clear evidence of this fact can be found in the Particle Data Book (PDG)~\cite{pdg}, where out of the six $D$ and five $D^*$ meson excited states listed (cataloging  separately the neutral and the charged ones), five $D$ and two $D^*$ states are omitted from the summary table due to controversy on their properties.  It is also interesting to notice that the energy region studied experimentally, 2400-2750 MeV, covers a range of only 350 MeV from the first to the last excited state, while the interaction of a $D$ or a $D^*$ meson, due to their heavy mass (around 2000 MeV), with a vector or few pseudoscalar mesons could give rise to a resonance with a mass close to 3000 MeV. Recently, a theoretical work along this line was made in Ref.~\cite{mknn}, where a system formed of a $D$ meson and the $f_0(980)$ resonance was investigated and a new $D$ meson state with mass around 2900 MeV was predicted. Similarly, with another recent study of the $D^*\rho$ system in s-wave using effective field theories~\cite{raquel} an effort has been made to shed some more light on the nature of two of the $D^*$ resonances listed by the PDG: $D^*_2(2460)$ and $D^*(2640)$. In Ref.~\cite{raquel}  the spin-parity of the former state  was confirmed to be $J^P=2^+$, and the spin and parity of the latter one was predicted to be $J^P=1^+$. In addition, a new state with mass close to 2600 MeV and spin 0, thus $J^P=0^+$, was predicted. 

In this talk I present the results obtained from our investigation of the $D^*\rho$ system using QCD sum rules to determine the mass of its possible resonances  and compare our results with the ones found in Ref.~\cite{raquel}.
\section{Framework}
The initial point to calculate the mass of the possible  $D^*\rho$ states with the QCD sum rule method consists in the evaluation of the two-point correlation function
\begin{align}
\Pi (q^2)=i\int d^4x\, e^{iqx}\langle 0|T\left[j(x)j^\dagger(0)\right]|0\rangle\label{corr},
\end{align}
where $q$ is the momentum flowing from $0$ to $\infty$, $T[\cdots]$ represents the $T$-ordered product and $j$ is the current associated to the $D^*\rho$ system.

Since we are working with two vector mesons, we can have total spin $S= 0, 1, 2$ and total isospin $I=1/2, 3/2$. Thus, we need to project the correlation function of Eq.~(\ref{corr}) on spin and isospin for a proper identification of the states.
For the isospin projection, we just need to consider the isospin relations
\begin{align}
|D^*\rho,I=1/2,I_z=1/2\rangle&=-\sqrt{\frac{2}{3}}|D^{*\,0}\rho^+\rangle+\sqrt{\frac{1}{3}}|D^{*\,+}\rho^0\rangle,\nonumber\\
|D^*\rho,I=3/2,I_z=3/2\rangle&=-|D^{*\,+}\rho^+\rangle.\label{isos}
\end{align}
where we adopt the phase convention $|\rho^+\rangle=-|1,1\rangle$ and $|D^{*\,0}\rangle=-|1/2,-1/2\rangle$. 

The simplest current we can use for the $D^*\rho$ system is of the form
\begin{align}
j_{\mu\nu}(x)=\left[\bar {q}^1_a(x)\gamma_\mu c_a(x)\right]\left[\bar{q}^2_b(x)\gamma_\nu q^3_b(x)\right],
\end{align}
with $q_1(x)$, $q_2(x)$ and $q_3(x)$ representing the fields of the light quarks $u$ or $d$, $c(x)$ is the field for the quark $c$, $a$ and $b$ are color indices and $\gamma$ represents the Dirac matrix. Considering Eq.~(\ref{isos}), the currents for the $D^*\rho$ system for the cases of total isospin $I=1/2$ and $3/2$ are thus,
\begin{align}
j^{1/2}_{\mu\nu}(x)&=-\sqrt{\frac{2}{3}}\left[(\bar{u}_a\gamma_\mu c_a)(\bar{d}_b\gamma_\nu u_b)-\frac{1}{2}(\bar{d}_a\gamma_\mu c_a)(\bar{u}_b\gamma_\nu u_b-\bar{d}_b\gamma_\nu d_b)\right],\label{cur}\\
j^{3/2}_{\mu\nu}(x)&=-(\bar{d}_a\gamma_\mu c_a)(\bar{d}_b\gamma_\nu u_b),\nonumber
\end{align}
where we have omitted the $x$ dependence of the quark fields for simplicity. 

For the spin projection, as shown in Refs.~\cite{markhem, khemmar}, we need to use the following projectors
\begin{align}
\mathcal{P}^{(0)}&=\frac{1}{3}\Delta^{\mu\nu}\Delta^{\alpha\beta},\nonumber\\
\mathcal{P}^{(1)}&=\frac{1}{2}\left(\Delta^{\mu\alpha}\Delta^{\nu\beta}-\Delta^{\mu\beta}\Delta^{\nu\alpha}\right),\label{DEproj}\\
\mathcal{P}^{(2)}&=\frac{1}{2}\left(\Delta^{\mu\alpha}\Delta^{\nu\beta}+\Delta^{\mu\beta}\Delta^{\nu\alpha}\right)-\frac{1}{3}\Delta^{\mu\nu}\Delta^{\alpha\beta},\nonumber
\end{align}
with
 \begin{align}
 \Delta_{\mu\nu}\equiv-g_{\mu\nu}+\frac{q_\mu q_\nu}{q^2},\label{Delta}\quad g_{\mu\nu}\equiv\textrm{metric tensor}.
 \end{align}
Equations (\ref{DEproj}) correspond to the covariant extrapolation of the nonrelativistic projectors found in Ref.~\cite{raquel} (for more details see Refs.~\cite{markhem, khemmar}), which implicitly assume that we have relative angular momentum $L=0$ for the two vector states. Thus we project the currents only on $J^P=$ $0^+$, $1^+$ and $2^+$.

In this way, the correlation function used to described the properties of the $D^*\rho$ system is given by
\begin{align}
 \Pi^{I,S}(q^2)=\mathcal{P}^{(S)}_\Delta\Pi^I_{\mu\nu,\alpha\beta}.\label{corr3},
 \end{align}
with
\begin{align}
\Pi^I_{\mu\nu,\alpha\beta}=i\int d^4x\, e^{iqx}\langle 0|T\left[j^I_{\mu\nu}(x){j^I_{\alpha\beta}}^\dagger(0)\right]|0\rangle.\label{corr2}
\end{align}

To evaluate Eq.~(\ref{corr2}) we need to rely on its dual nature: for large momentum, i.e., short distances, the correlation function represents a quark-antiquark fluctuation and can be calculated using perturbative QCD. At large distances, or, equivalently, small momentum, the currents $j^\dagger$ and $j$ of Eq.~(\ref{corr}) can be interpreted as operators of creation and annihilation of hadrons with the same quantum numbers as the ones of the current $j$. In this case, the correlation function is obtained by inserting a complete set of states with the same quantum numbers as those of the current under consideration. The first way of evaluating the correlation function is called ``OPE description'' (Operator Product Expansion), while the second receive the name of ``Phenomenological description.''

In the OPE description of the correlation function, Eq.~(\ref{corr3}) is written as
\begin{align}
 \Pi^{I,S}_{\textrm{OPE}}(q^2)=\int_{m^2_c}^\infty ds\frac{\rho^{I,S}_{\textrm{OPE}}(s)}{s-q^2},\label{corrope}
\end{align}
with $\rho^{I,S}_{\textrm{OPE}}(s)$ being the spectral density, which is related to the imaginary part of the correlation function through $\pi \rho^{I,S}_{\textrm{OPE}}=\textrm{Im}\left[ \Pi^{I,S}_{\textrm{OPE}}\right]$. In this study of the $D^*\rho$ system we have considered condensates up to dimension 7 in the OPE description. In this way:
\begin{align}
\rho^{I,S}_{\textrm{OPE}}&=\rho^{I,S}_{\textrm{pert}}+\rho^{I,S}_{\langle\bar q q\rangle}+\rho^{I,S}_{\langle g^2 G^2\rangle}+\rho^{I,S}_{\langle\bar q g\sigma G q\rangle}+\rho^{I,S}_{{\langle\bar q q\rangle}^2}+\rho^{I,S}_{\langle g^3 G^3\rangle}+\rho^{I,S}_{\langle\bar q q\rangle\langle g^2 G^2\rangle}.\label{rho}
\end{align}
The first term in Eq.~(\ref{rho}) or perturbative term corresponds to dimension 0 in the expansion. The next term, i.e., the quark condensate contribution ($\langle\bar q q\rangle$) has dimension 3. The two gluon condensate contribution ($\langle g^2 G^2\rangle$) corresponds to dimension 4 in the expansion. The mixed condensate or $\langle\bar q g\sigma G q\rangle$ term contributes with dimension 5. The two quark and the three gluon condensates (${\langle\bar q q\rangle}^2$ and $\langle g^3 G^3\rangle$, respectively) have dimension 6 and, finally, the contribution associated to the condensate 
$\langle\bar q q\rangle\langle g^2 G^2\rangle$ has dimension 7. The result for each of the terms of Eq.~(\ref{rho}) for the different isospins and spins can be found in the Appendix of Ref.~\cite{markhem} and the value of the different condensates is listed in Table~\ref{table}.

\begin{table}[h!]
\caption{Values of the different parameters required for numerical 
calculations of the correlation function given by Eq.~(\ref{corr3}) 
(see Ref.~\cite{narison}).}\label{table}
\begin{tabular}{cc}
Parameters & Values\\
\hline
$m_c$& $1.23 \pm 0.05$ GeV\\
$\langle \bar{q} q \rangle$ & $-(0.23 \pm 0.03)^3$ GeV$^3$\\
$\langle g^2 G^2 \rangle$ & $(0.88\pm 0.25)$ GeV$^4$\\
$\langle g^3 G^3 \rangle$ & $(0.58\pm 0.18)$ GeV$^6$\\
$\langle \bar{q} \sigma \cdot G q \rangle$& 0.8$\langle \bar{q} q \rangle$ GeV$^2$\\
\end{tabular}
\end{table}

In the phenomenological description,  the correlation function for the $D^*\rho$ system can be written in terms of a spectral density as
\begin{align}
 \Pi^{I,S}_{\textrm{phenom}}(q^2)=\int_{m^2_c}^\infty ds\frac{\rho^{I,S}_{\textrm{phenom}}(s)}{s-q^2}.\label{corr5}
\end{align}
All hadrons with the same quantum numbers as the ones associated to the current $j$ of Eq.~(\ref{corr}) contribute to the density of Eq.~(\ref{corr5}). Therefore, to extract information about the states we are interested in,
the spectral density $\rho^{I,S}_{\textrm{phenom}}$ needs to be parametrized conveniently. Commonly, the density of Eq.~(\ref{corr5}) is expressed as a sum of a narrow, sharp state, which represents the one we are looking for, and a smooth continuum~\cite{svz,io,narison,marina}
\begin{align}
\rho^{I,S}_{\textrm{phenom}}(s)={\lambda^2_{I,S}}\delta(s-m_{I,S}^2)+\rho^{I,S}_{\textrm{cont}}(s),
\end{align}
with $\lambda^2_{I,S}$ the coupling of the current to the state we are interested in and $m_{I,S}$ its mass. The density related to the continuum of states is assumed to vanish below a certain value of $s$, $s_0$, called continuum threshold. Above this threshold it is common to consider the ansatz~\cite{svz,io,narison,marina}
\begin{align}
\rho_{\textrm{cont}}(s)=\rho^{I,S}_{\textrm{OPE}}(s)\Theta(s-s^{I,S}_0).\label{ansatz}
\end{align}
Therefore, using this parametrization of the spectral density of Eq.~(\ref{corr5}), the correlation function in the phenomenological description adopts the form
\begin{align}
 \Pi^{I,S}_{\textrm{phenom}}(q^2)=\frac{\lambda^2_{I,S}}{m_{I,S}^2-q^2}+\int_{s^{I,S}_0}^\infty ds\,\frac{\rho^{I,S}_{\textrm{OPE}}(s)}{s-q^2}.\label{corrphen}
\end{align}
Ideally, the result from the evaluation of Eqs.~(\ref{corrope}) and (\ref{corrphen}) should be the same at some range of $q^2$ at which we could just directly equate both expressions.
However, this is not completely true: the correlation function in Eq.~(\ref{corrope}) suffers from divergent contributions originated from long range interactions,
while the one in Eq.~(\ref{corrphen}) contains contributions arising from the continuum. These problems can be solved by applying the Borel transformation
to both correlation functions and then equating them, which gives the relation
\begin{align}
\lambda^2_{I,S}e^{-m_{I,S}^2/M^2}=\int_{m^2_c}^{s^{I,S}_0}ds\,\rho^{I,S}_{\textrm{OPE}}(s)e^{-s/M^2},\label{mrule}
\end{align}
where $M$ is the Borel mass parameter. Calculating the derivative of Eq.~(\ref{mrule}) with respect to $-M^{-2}$ and dividing the resulting expression by Eq.~(\ref{mrule}), we can determine the mass as
\begin{align}
m_{I,S}^2=\frac{\int_{m^2_c}^{s^{I,S}_0}ds\,s \rho^{I,S}_{\textrm{OPE}}(s)e^{-s/M^2}}{\int_{m^2_c}^{s^{I,S}_0}ds\,\rho^{I,S}_{\textrm{OPE}}(s)e^{-s/M^2}}.\label{mass}
\end{align}
Once the mass is obtained, we can calculate the coupling $\lambda_{I,S}$ through Eq.~(\ref{mrule})
\begin{align}
\lambda^2_{I,S}=\frac{\int_{m^2_c}^{s^{I,S}_0}ds\,\rho^{I,S}_{\textrm{OPE}}(s)e^{-s/M^2}}{e^{-m_{I,S}^2/M^2}}\label{lambda}.
\end{align}

As can be seen in Eqs.~(\ref{mass}) and (\ref{lambda}), the masses and couplings are functions of the Borel mass $M$. In an idealistic situation, the results should not depend on the value of the Borel mass used to determine them. But, in a realistic case, this is not completely true and one relies on the existence of a range of Borel masses (or Borel ``window'') in which the results obtained for the mass and the coupling can be relied upon. The determination of this Borel window is based on the following constraints: (1) In the phenomenological description, the contribution to the correlation function arising from the pole term, which represents the state in which we are interested in, should dominate over the contribution from the continuum of states with the same quantum numbers. This condition fixes the maximum value for the Borel mass, $M_\textrm{max}$, at which the results obtained for the mass and coupling are meaningful (2). The other constraint is to guarantee the convergence of the OPE expansion. This fixes the minimum Borel mass at which the sum rule is reliable, $M_\textrm{min}$.~Here, in this manuscript, if $n$ is the maximum dimension taken into account in the OPE expansion, we consider that the right side of Eq.~(\ref{mrule}) converges when the relative contribution associated to the condensate of dimension $n-1$, i.e, the result of dividing the contribution to the integral in Eq.~(\ref{mrule}) of all terms of the series up to dimension $n-1$ by the contribution to the integral associated to all the terms,  differs, in modulus, by no more than 10-25 \% (depending on the case) from the relative contribution to the condensate of dimension n. 

At the same time, the Borel mass window depends on the continuum threshold, $s^{I,S}_0$. This continuum threshold is a parameter of the model, but its value is not completely arbitrary. As can be seen in Eq.~(\ref{ansatz}), $\sqrt{s^{I,S}_0}$ is related to the onset of the continuum in the current $j$ under consideration and a reasonable value is about 450-500 MeV above the mass of the hadron we are looking for~\cite{marina}.  For this case, where we are searching for possible $D^*\rho$ molecular states of masses 2.450-2.600 GeV, we will use values for $\sqrt{s^{I,S}_0}\sim 3.00-3.15$ GeV.

\section{Results}
\begin{figure}
\includegraphics[width=0.4\textwidth]{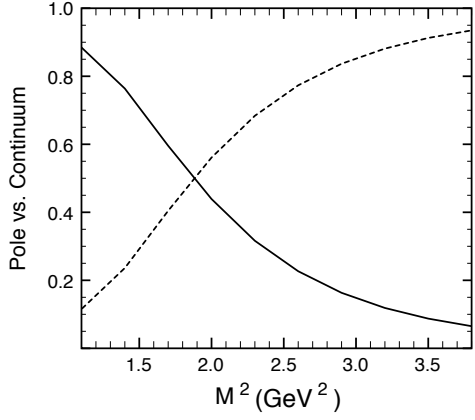}
\caption{Contributions of the pole (solid line) and the continuum (dashed line) terms (weighted by their sum) to the correlation function in the phenomenological description for the case $I=1/2$, $S=2$ as function of the squared Borel mass. These results are obtained for a value of the continuum threshold of 3.15 GeV and using the central values of Table~\ref{table} for the quark masses and condensates.}\label{polecont}
\end{figure}
\begin{figure}
\includegraphics[width=0.4\textwidth]{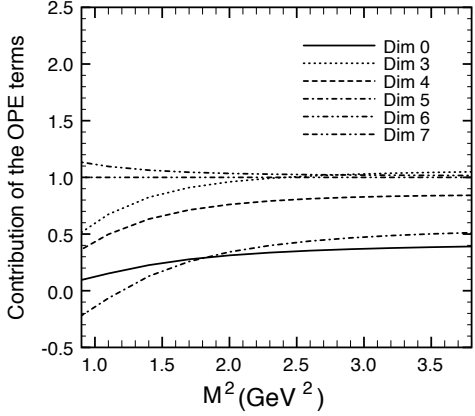}
\caption{Relative contributions of the different terms of the series for the OPE description of the correlation function for the case $I=1/2$ and $S=2$ as a function of the squared Borel mass. These results are obtained for a value of the continuum threshold of 3.15 GeV and using the central values of Table~\ref{table} for the quark masses and condensates.}\label{OPE}
\end{figure}
\begin{figure}
\includegraphics[width=0.4\textwidth]{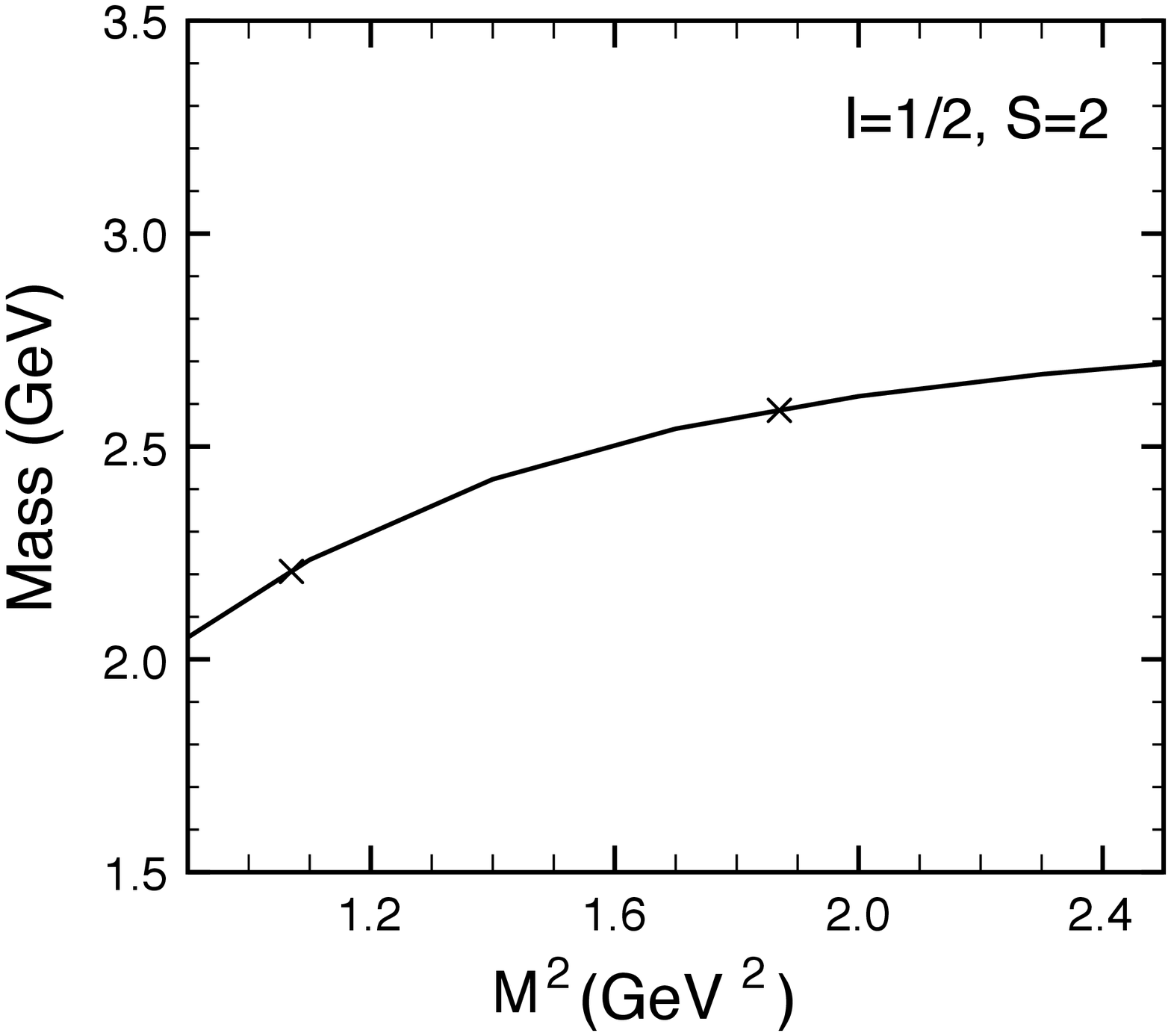}
\caption{Mass sum rule for the case $I=1/2$ and $S=2$ as a function of the squared Borel mass. The crosses in the figure indicate the Borel mass window. This result is found using a value of the continuum threshold of 3.15 GeV and considering the central values of Table~\ref{table} for the quark masses and condensates.}\label{mass2}
\end{figure}

In Fig.~\ref{polecont} we show as a function of the Borel mass and for the phenomenological description of the correlation function the contributions of the pole term and the continuum, weighted by their sum, for the case $I=1/2$, $S=2$, and a value of $\sqrt{s^{1/2,2}_0}=3.15$ GeV. As can be seen, the pole term dominates over the continuum of states for a value of the squared Borel mass of 1.87 GeV$^{2}$, thus, $M^2_\textrm{max}=1.87$ GeV$^{2}$. Similarly, for the OPE description of the correlation function, we show in Fig.~\ref{OPE} the result of the study of the convergence of the OPE series considering the central values of the quark masses and condensates of Table~\ref{table}. We find a convergence of the OPE series for a value of the squared Borel mass above 1.07 GeV$^2$, thus, $M^2_\textrm{min}=1.07$ GeV$^{2}$. Once a Borel window is found, we calculate the mass resulting from Eq.~(\ref{mass}), and obtain, as can be seen in Fig.~\ref{mass2}, that the mass sum rule is stable and gives as a result
\begin{align}
m_{1/2,2}=(2.428\pm0.151)\, \textrm{GeV}.\label{mres}
\end{align}

The value shown in Eq.~(\ref{mres}) is obtained by averaging the mass over the Borel window and by calculating the standard deviation to determine the error. Similarly, using Eq.~(\ref{lambda}), we can get the coupling of the state to the current used and we find
\begin{align}
\lambda_{1/2,2}=(8.10\pm 2.00)\times 10^{-3}\, \textrm{GeV}^5.\label{lres}
\end{align}

To estimate the uncertainty of the  results in Eqs.~(\ref{mres}) and (\ref{lres}), we consider the change found in the mass of the state while varying $\sqrt{s^{1/2,2}_0}$ in the interval 3.00-3.15 GeV and the quark masses and condensates within the error related to them (shown in Table \ref{table}). Taking into account all these sources of errors, averaging over the results found in the different Borel windows and calculating the standard deviation, we obtain
\begin{align}
m_{1/2,2}&=(2.439\pm0.119)\, \textrm{GeV},\nonumber\\
\lambda_{1/2,2}&=(8.14\pm 1.61 )\times 10^{-3}\, \textrm{GeV}^5.\label{m2}
\end{align}

Analogously, we repeat this process for the cases $I=1/2, 3/2$ and spin $0$ and $1$, respectively. In all cases, we find a valid Borel window considering the dominance of the pole term over the continuum in the phenomenological description and the convergence of the series in the OPE description. We also change the value of the continuum threshold in the
 range $\sqrt{s^{1/2,0}_0}=\sqrt{s^{1/2,1}_0}\sim$ 3.00-3.15 GeV, and the quark masses and condensates in the range shown in Table~\ref{table}.  Averaging over the results found in the different Borel windows and calculating the standard deviation we obtain:
\begin{align}
m_{1/2,0}&=(2.500\pm0.067)\, \textrm{GeV},\nonumber\\
\lambda_{1/2,0}&=(3.63\pm 0.39 )\times 10^{-3}\, \textrm{GeV}^5,\nonumber\\
m_{1/2,1}&=(2.523\pm0.060)\, \textrm{GeV},\label{m01}\\
\lambda_{1/2,1}&=(6.51\pm 0.61)\times 10^{-3}\, \textrm{GeV}^5.\nonumber\\
m_{3/2,0}&=(2.467\pm0.082)\, \textrm{GeV},\nonumber\\
\lambda_{3/2,0}&=(3.48\pm 0.48 )\times 10^{-3}\, \textrm{GeV}^5,\nonumber\\
m_{3/2,1}&=(2.420\pm 0.128)\, \textrm{GeV},\nonumber\\
\lambda_{3/2,1}&=(5.22\pm 1.12)\times 10^{-3}\, \textrm{GeV}^5,\nonumber\\
m_{3/2,2}&=(2.550\pm 0.056)\, \textrm{GeV},\nonumber\\
\lambda_{3/2,2}&=(6.60\pm 0.62)\times 10^{-3}\, \textrm{GeV}^5.\nonumber
\end{align}

Therefore, we can conclude that three states are found for the case $I=1/2$, each with a different spin $S=0$, $1$, and $2$, with the corresponding masses and couplings given by Eqs. (\ref{m2}) and (\ref{m01}).  It is interesting to note that
the coupling of the spin 2 state to the current is bigger than the respective ones for spin 0 and 1.  These results are in striking agreement with the ones obtained in Ref.~\cite{raquel}. The state found in this manuscript with spin 2 can be associated with the $D^*_2(2450)$ listed by the PDG. For the states with spin 0 and 1, there is only one candidate listed by the PDG and that is $D^*(2640)$. However, nothing is known about the spin and parity of this state. In Ref.~\cite{raquel}, the widths related to the resonances found were calculated, obtaining  a width of around 40 MeV for the state with spin 2, 60 MeV for the state with spin 0 and practically zero width for the state with spin 1. Since the width listed by the PDG for $D^*(2640)$ is $\Gamma < 15$ MeV, the authors of Ref.~\cite{raquel} associated the state with spin 1 to $D^*(2640)$ and predicted the existence of a state with spin 0 and a similar mass, but with a much larger width. In this manuscript, we have calculated the masses and couplings for the different states. To make a proper identification of $D^*(2640)$ with one of the states with spin 0 and 1, a QCD sum rule calculation to determine the width of each of the states might be helpful. However, this is beyond the scope of the present manuscript. Thus, with the information at hand, we can only confirm the existence of two nearly degenerate states with masses around $2500\pm60$ MeV and spins 0 and 1, respectively, one of which can probably be related to $D^*(2640)$.

For the case $I=3/2$, we find also three states, one for each spin case. The study made by the authors of Ref.~\cite{raquel} disfavor the formation of molecular resonances in the $D^*\rho$ system for total isospin $3/2$. This is due to the fact that the kernel which enters in the resolution of the Bethe-Salpeter equation,
is repulsive for this particular isospin and for the three possible spins. Actually, this result found for the $D^*\rho$ system is similar to the one found in the study of other hadronic systems using an approach with the same spirit of Ref.~\cite{raquel}. Thus, in general, in such approaches whenever the system can generate a resonance whose quantum numbers can not be explained in terms of the constituent quark model, a repulsive interaction is found,  preventing the formation of such a state. The experimental search of these states will be definitely very useful to clarify the existence of possible exotic molecular hadron resonances formed due to the interaction of a $D^*$ and a $\rho$ and, at the same time, a test of the QCD sum rule approach used in this manuscript.
\section{Summary}
Using an approach based on QCD sum rules, we have studied the interaction of a $D^*$ and a $\rho$ mesons and investigated the existence of resonances for the different isospins and spins.
For the isospin $1/2$ case, we have found three states with masses $2.500\pm0.067$ GeV, $2.523\pm0.060$ GeV and $2.439\pm0.119$ GeV and spin 0, 1 and 2, respectively. The state with spin 2 can be associated with the meson $D^*_2(2450)$ listed by the PDG.  One of the other two resonances with spin $0$ or $1$ can be related to the meson $D^*(2640)$ of the PDG, whose spin-parity is unknown. For the case of isospin $3/2$, we have obtained three states with masses $2.467\pm0.082$ GeV, $2.420\pm0.128$ GeV, $2.550\pm0.056$ GeV and spins 0, 1, and 2, respectively. These states can be considered exotic in the sense that their quantum numbers can not be obtained from a quark plus antiquark configuration. While the results found for the isospin 1/2 case are in good agreement with the ones obtained by a previous study (see Ref.~\cite{raquel}) of the same system with a different approach, the results for the isospin 3/2 sector are different. The experimental studies of this system will be very relevant to confirm the results found by the model of Ref.~\cite{raquel} and the one of this manuscript in the isospin $1/2$ sector and the possible existence of $D^*\rho$ molecular resonances with isospin $3/2$ and spin 0, 1, and 2.


\begin{theacknowledgments}
 We thank professors Altu\u{g} \"Ozpineci and Juan Nieves for very useful discussions and for a careful reading of the manuscript. The authors would like to thank the Brazilian funding agencies FAPESP and CNPq for the financial support. This work is partly supported by the Spanish Ministerio de Economia y Competividad and European FEDER fund under the contract number FIS2011-28853-C02-01 and the Generalitat Valenciana in the program Prometeo, 2009/090. E. Oset acknowledges the support of the European Community-Research Infrastructure Integrating Activity Study of Strongly Interacting Matter (acronym HadronPhysics3, Grant Agreement n. 283286) under the Seventh Framework Programme of EU.
 \end{theacknowledgments}


\bibliographystyle{aipproc}

\begin{thebibliography}{99}

\bibitem{markI}
 G.~Goldhaber, F.~Pierre, G.~S.~Abrams, M.~S.~Alam, A.~Boyarski, M.~Breidenbach, W.~C.~Carithers and W.~Chinowsky {\it et al.},
  \emph{Phys.\ Rev.\ Lett.}  {\bf 37}, 255 (1976).
  
  G.~Goldhaber, J.~Wiss, G.~S.~Abrams, M.~S.~Alam, A.~Boyarski, M.~Breidenbach, W.~Chinowsky and J.~Dorfan {\it et al.},
  Phys.\ Lett.\ B {\bf 69}, 503 (1977).
  
  \bibitem{cleo}
  P.~Avery {\it et al.}  [CLEO Collaboration],
  Phys.\ Rev.\ D {\bf 41}, 774 (1990).
  
  \bibitem{belle}
  K.~Abe {\it et al.}  [Belle Collaboration],
  Phys.\ Rev.\ D {\bf 69}, 112002 (2004).
  
  \bibitem{babar}
  P.~del Amo Sanchez {\it et al.}  [BaBar Collaboration],
  Phys.\ Rev.\ D {\bf 82}, 111101 (2010).
  
 \bibitem{pdg}
 J. Beringer et al. (Particle Data Group), Phys. Rev. D {\bf 86}, 010001 (2012).
 
 \bibitem{mknn} 
  A.~Martinez Torres, K.~P.~Khemchandani, M.~Nielsen and F.~S.~Navarra,
  Phys.\ Rev.\ D {\bf 87}, 034025 (2013)
  
 \bibitem{raquel}
 R.~Molina, H.~Nagahiro, A.~Hosaka and E.~Oset,
  Phys.\ Rev.\ D {\bf 80}, 014025 (2009).
  \bibitem{markhem} 
  A.~Martinez Torres, K.~P.~Khemchandani, M.~Nielsen, F.~S.~Navarra and E.~Oset,
  arXiv:1307.1724 [nucl-th], accepted for publication in Phys. Rev. D.
    \bibitem{khemmar} 
  K.~P.~Khemchandani, A.~Martinez Torres, M.~Nielsen and F.~S.~Navarra,
  arXiv:1310.0862 [hep-ph].
\bibitem{svz} M.A. Shifman, A.I. and Vainshtein and V.I. Zakharov,
Nucl. Phys. B {\bf 147}, 385 (1979).

\bibitem{io}
B.~L. Ioffe, Nucl.\ Phys.\ B {\bf 188}, 317 (1981); B {\bf 191}, 591(E) (1981).

  \bibitem{narison} S.~Narison, Phys.\ Lett.\ B {\bf 216}, 191 (1989).
  
 \bibitem{marina} 
   M.~Nielsen, F.~S.~Navarra and S.~H.~Lee,
  Phys.\ Rept.\  {\bf 497}, 41 (2010).
  


\end{thebibliography}

\end{document}